\documentclass[12pt,preprint]{aastex}

\newcommand{\etal}{et~al.\ }

\newcommand{\CIV}{C~{\sc iv}}

\newcommand{\kms}{\hbox{km~s$^{-1}$}}
\newcommand{\cmsq}{\hbox{cm$^{-2}$}}

\newcommand{\nh}{\hbox{${N}_{\rm H}$}}

\begin{document}

\def\sarc{$^{\prime\prime}\!\!.$}
\def\arcsec{$^{\prime\prime}$}
\def\arcmin{$^{\prime}$}
\def\degr{$^{\circ}$}
\def\seco{$^{\rm s}\!\!.$}
\def\ls{\lower 2pt \hbox{$\;\scriptscriptstyle \buildrel<\over\sim\;$}} 
\def\gs{\lower 2pt \hbox{$\;\scriptscriptstyle \buildrel>\over\sim\;$}} 
 
\title{2MASS Reveals a Large Intrinsic Fraction of BALQSOs}

\author{Xinyu Dai, Francesco Shankar, and Gregory R. Sivakoff}

\altaffiltext{1}{Department of Astronomy,
 The Ohio State University, Columbus, OH 43210,
 xinyu@astronomy.ohio-state.edu, shankar@astronomy.ohio-state.edu, sivakoff@astronomy.ohio-state.edu}

\begin{abstract}
The intrinsic fraction of broad absorption line quasars (BALQSOs) is important
in constraining geometric and evolutionary models of quasars.
We present the fraction of BALQSOs in 2MASS detected quasars within the SDSS DR3
sample in the redshift range of $1.7 \le z \le 4.38$.
The fraction of BALQSOs is $40.4^{+3.4}_{-3.3}$\% in the 2MASS 99\% database
$K_s$ band completeness sample, and $38.5^{+1.7}_{-1.7}$\% in
the larger 2MASS sample extending below the completeness limit.
These fractions are significantly higher than the 26\% reported in the optical
bands for the same parent sample.
We also present the fraction of BALQSOs as functions of apparent magnitudes,
absolute magnitudes, and redshift in the 2MASS and SDSS bands.
The 2MASS fractions are consistently higher than the SDSS fractions in every comparison,
and the BALQSO fractions steadily increase with wavelength from the
SDSS $u$ to the 2MASS $K_s$ bands.
Furthermore, the $i - K_s$ color distributions of BALQSOs and non-BALQSOs
indicate that BALQSOs are redder than non-BALQSOs, with a K-S test probability
of $2\times10^{-12}$.
These results are consistent with the spectral difference between BALQSOs and non-BALQSOs 
including both the absorption troughs and dust extinction in BALQSOs, which
leads to significant selection biases against BALQSOs in the optical bands.
Using a simple simulation incorporating the luminosity function of quasars
and the amount of obscuration for BALQSOs, we simultaneously fit 
the BALQSO fractions 
in the SDSS and 2MASS bands. 
We obtain a true BALQSO fraction of $43\pm2$\% for luminous quasars ($M_{K_s} \lesssim -30.1$~mag).

\end{abstract}


\section{Introduction}
The broad absorption line quasars (BALQSO) are a sub-sample of quasars
exhibiting blue-shifted rest-frame ultra-violet absorption troughs (e.g.,
Weymann et al. 1991).
In X-rays, typical \nh\ absorption columns of $10^{22-24}\cmsq$
(e.g., Gallagher et al. 1999, 2002; Green et al. 2001; Chartas et al. 2001;
Grupe et al. 2003) explain the X-ray weakness of the population.
The BALQSOs are important in understanding the properties of
quasars.
In geometric models of quasars, BALQSOs are quasars viewed at large inclination
angles close to the equatorial plane (e.g., Weymann et al. 1991; Ogle et al. 
1999; Schmidt \& Hines 1999; Hall et al. 2002) or the polar direction (Zhou et al. 2006).
Other models place
BALQSOs at the early stages of quasar evolution (e.g., Hazard et al. 1984;
Surdej \& Hutsemekers 1987; Boroson \& Meyers 1992; Becker et al. 2000).

Under the recent paradigm of the co-evolution of AGN and the host galaxies, 
BALQSOs are expected to be a manifestation of the AGN kinetic feedback, an energetic output that many recent theoretical models (e.g.,
Granato et al. 2004) of galaxy evolution advocate to reproduce several observables such as the galaxy stellar mass function. 
In this scenario the majority of luminous AGNs must have undergone a wind phase and therefore a BAL phase. 
On the other hand, optical surveys have always detected a small fraction of BALQSOs within their sample. 
Several studies of optically selected quasar samples indicate that the fraction of BALQSOs, $f_{BAL} = {N_{BAL}}/{N_{Total}}$
is in the range of 10-22\% (e.g., Weymann et al. 1991; Tolea et al. 2002; 
Hewett \& Foltz 2003; Reichard et al. 2003, R03 hereafter).
In the radio bands, recent FIRST survey
results show that the fraction of BALQSOs in radio selected quasars is about
14--18\% (Becker et al. 2000).
In the infrared bands, there were hints of larger BALQSO fractions (Voit et
al. 1993; Lipari 1994; Egami et al. 1996), but the existing results are difficult 
to interpret due to the small sample size.
Special geometries of the outflows or short duty cycles have then been advocated to explain the deviation 
from expectations.
However, there are other observational results suggesting larger fractions of BALQSOs.
Notably, in a gravitationally lensed quasar sample, Chartas (2000) found a
BALQSO fraction of 35\%, however, the sample size was small.
Recently, Trump et al. (2006, T06 hereafter) identified more than 4,000 BALQSOs
in the large SDSS DR3 quasar sample (Schneider et al. 2005), and found a BALQSO
fraction of 26\%.

One important difficulty in estimation of the true BALQSO fraction is the selection bias
caused by the UV spectral difference between BALQSOs and non-BALQSOs, which includes
 both absorption troughs and continuum differences 
(Sprayberry \& Foltz 1992; R03).
However, direct modelings of this selection bias lead to different correction factors 
(e.g., Hewett \& Foltz 2003; R03), though the issue is further complicated by different
sample selections.
Alternatively, we can study the fraction of BALQSOs in a spectral regime that is not
significantly affected by these biases to obtain the true fraction of BALQSOs.
In this paper, we examine the population of BALQSOs in
the near infrared by analyzing the large sample of SDSS-BALQSOs (T06) combining
 2MASS (Skrutskie et al. 2006) data.
We note that the BALQSO fraction also depends on the definition of BALQSOs used by
different studies, and we adopt the definition used by T06.

We assume that $H_0 = 70~\rm{km~s^{-1}~Mpc^{-1}}$, $\Omega_{\rm m} = 0.3$, 
and $\Omega_{\Lambda}= 0.7$ throughout the paper.

\section{Sample Selection}

\begin{figure}
\plotone{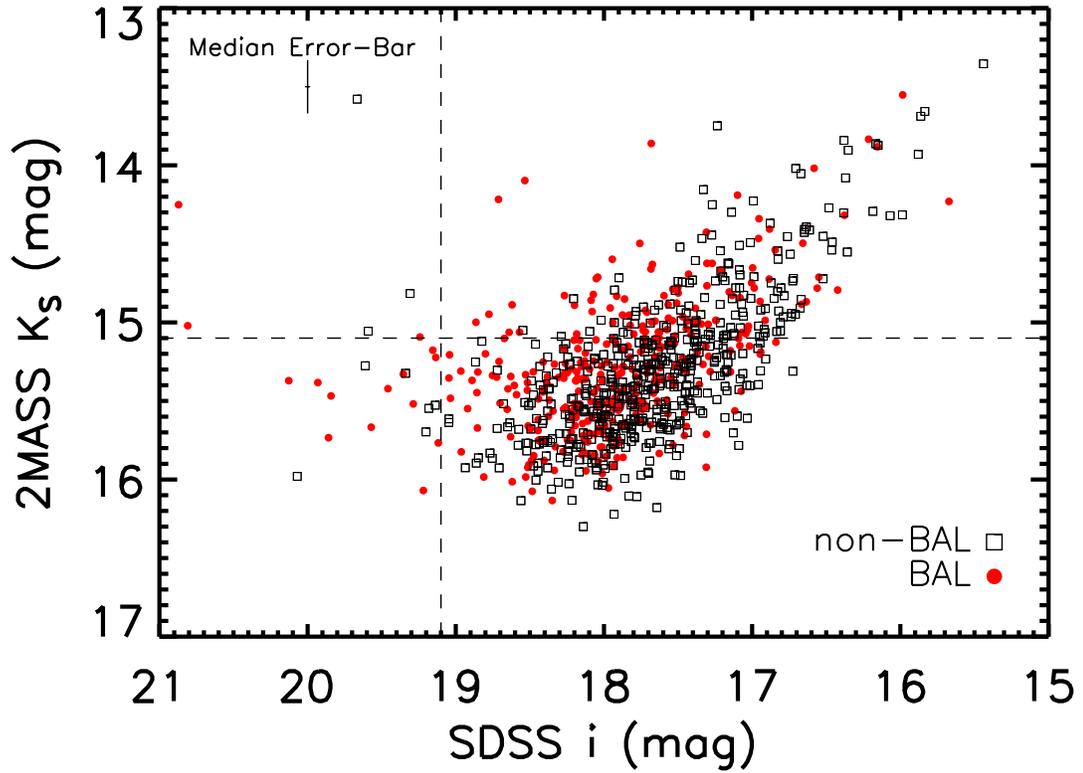}
\caption{2MASS $K_s$ mag versus SDSS $i$ mag for BALQSOs (red circles) and
non-BALQSOs (black squares) detected in all of the $J$, $H$, and $K_s$ bands in
the redshift range of $1.7 \le z \le 4.38$. The dashed line show the 99\%
completeness $K_s$ limit 
for the 2MASS database entries and the SDSS quasar selection limit of $i<19.1$. The
sample is limited by the 2MASS flux limit, rather than the SDSS quasar selection limit.\label{fig:ik}}
\end{figure}

We started from the SDSS DR3 quasar catalog (Schneider et al. 2005) and the
SDSS-BALQSO catalog (T06). In particular, we targeted the redshift
range of $1.7 \le z \le 4.38$, where the majority of BALQSOs are identified
with \CIV\ absorption in the observed-frame optical band pass. 
We matched the quasars of both BALQSOs and non-BALQSOs
(2\arcsec) to entries from the full 2MASS release. We note that not all database
entries from the full 2MASS release in the 2MASS All-Sky Point Source Catalog
(PSC) satisfy the conservative requirements to be part of the official PSC. 
Therefore, the completeness of the database entries extend below the 
official PSC completeness limits\footnote{See
\url{http://www.ipac.caltech.edu/2mass/releases/allsky/doc/sec6\_5a1.html}.}. 
The 99\% completeness levels of the database are $J=16.1$, $H=15.5$, and
$K_{s}=15.1$~mag. To ensure that matched database entries represent detections, we
required an in-band detection (rd\_flg != 0), that there was no confusion,
contamination (cc\_flg == 0), or blending (bl\_flg $\le$ 1) in the source, and that no source was near an extended galaxy (gal\_contam == 0 and ext\_key is
null). Fig.~\ref{fig:ik} shows the $K_s$ mag versus $i$ mag distributions for
BALQSOs and non-BALQSOs.
The quasars are concentrated in a linear relation between the $i$ and $K_s$
magnitudes with scatter. 
The near
infrared quasar sample is limited by the 2MASS flux limits, rather than the SDSS
quasar selection. The first flux limit for SDSS quasar selection is at $i<19.1$,
where quasars are selected in the $ugri$ color cube (Schneider et al. 2005), and
the 2MASS sample fails to detect the majority of quasars at this SDSS $i$
magnitude limit. About 5.2\% of SDSS quasars are detected in 2MASS. 
This is expected since the 2MASS is a significantly shallower 
survey than the SDSS survey.
We define the ``2MASS sample'' as quasars detected in all of the $J$, $H$, and $K_s$ bands, 
and the ``$K_s$ complete sample'' as quasars with $K_s < 15.1$~mag.
In \S5, we use quasar luminosity functions to model BALQSO fractions above selected absolute magnitude
limits.
Since 2MASS cannot detect quasars less luminous than the above limits at high redshifts, 
we also present results for a ``narrower redshift'' sample, $1.7 < z < 2.5$.
Our results are not 
significantly affected by the choice of sample selection.

\section{The BALQSO Fraction in 2MASS}

We found 884 quasars in the 2MASS sample with 340 BALQSOs and 544 non-BALQSOs.
In the $K_s$ complete sample, we found 245, 99, and 146
quasars for total, BALQSOs, and non-BALQSOs, respectively.
The BALQSO fraction is $40.4^{+3.4}_{-3.3}$\% in the $K_s$ complete sample and
$38.5^{+1.7}_{-1.7}$\% in the 2MASS sample. 
We also tested $J$ and $H$ band
complete samples, obtaining consistent results, $37.7^{+4.2}_{-4.1}$\% and
$35.6^{+4.6}_{-4.4}$\%, respectively. These BALQSO fractions from 2MASS bands
are significantly higher than the 26\% in the optical bands (T06). For example,
the difference between the BALQSO fraction in optical and the $K_s$ complete
sample is 4.4$\sigma$. This result is interesting, as we are analyzing
the same parent BALQSO catalog used in the optical study. It indicates that
there may be significant selection bias against BALQSOs in the optical
photometric surveys. As a corollary the near infrared fraction obtained in this paper
(35-40\%) represents the true fraction of BALQSOs. Alternatively, BALQSOs may be
intrinsically brighter in the near infrared bands or a combination of both effects
is possible.  We examine the fraction of BALQSOs as functions of apparent magnitude, 
absolute magnitude, and redshift to investigate the cause for the 
increase of the BALQSO fractions in the 2MASS bands.

\begin{figure}
\epsscale{0.8}
\plotone{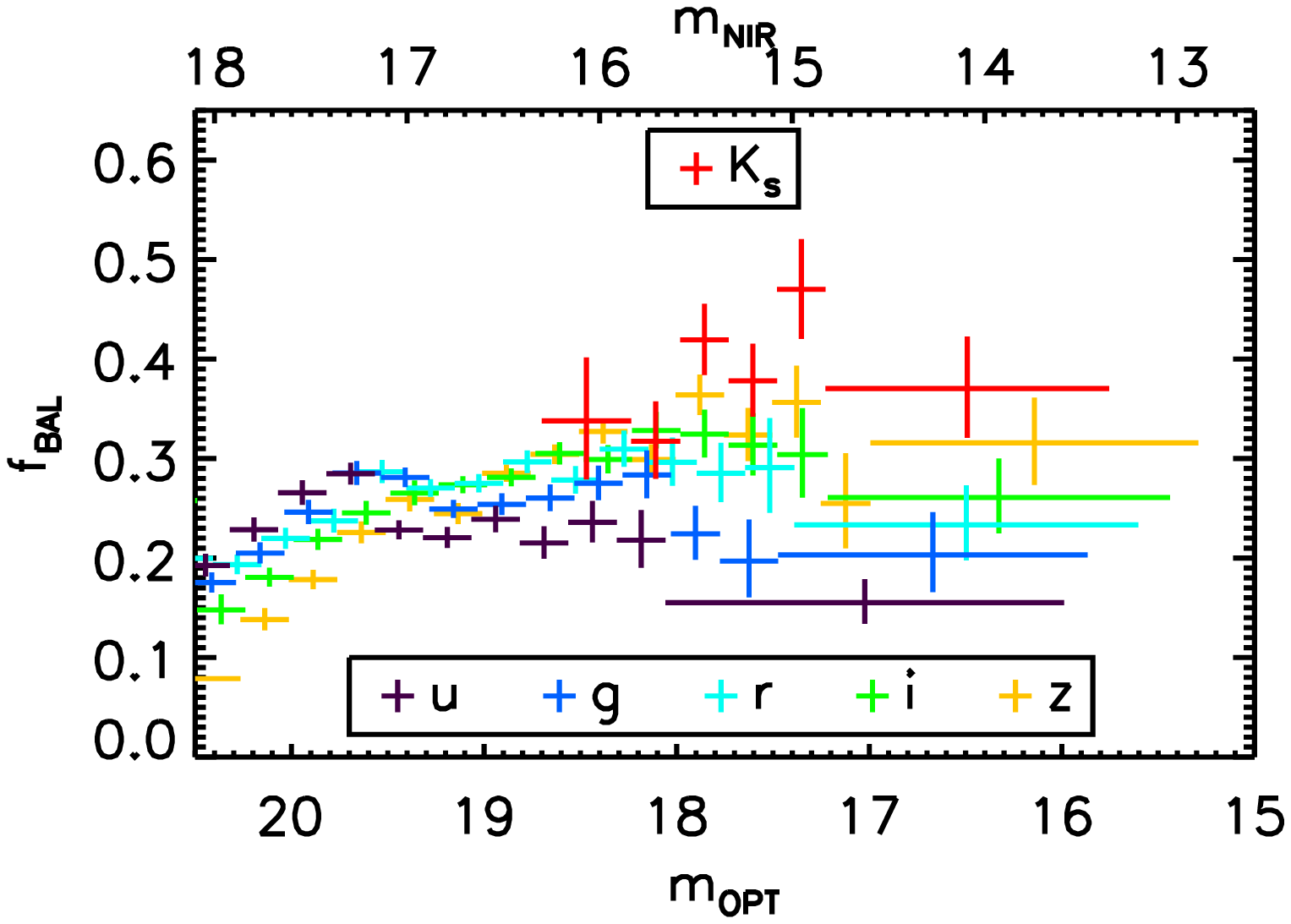}\\
\plotone{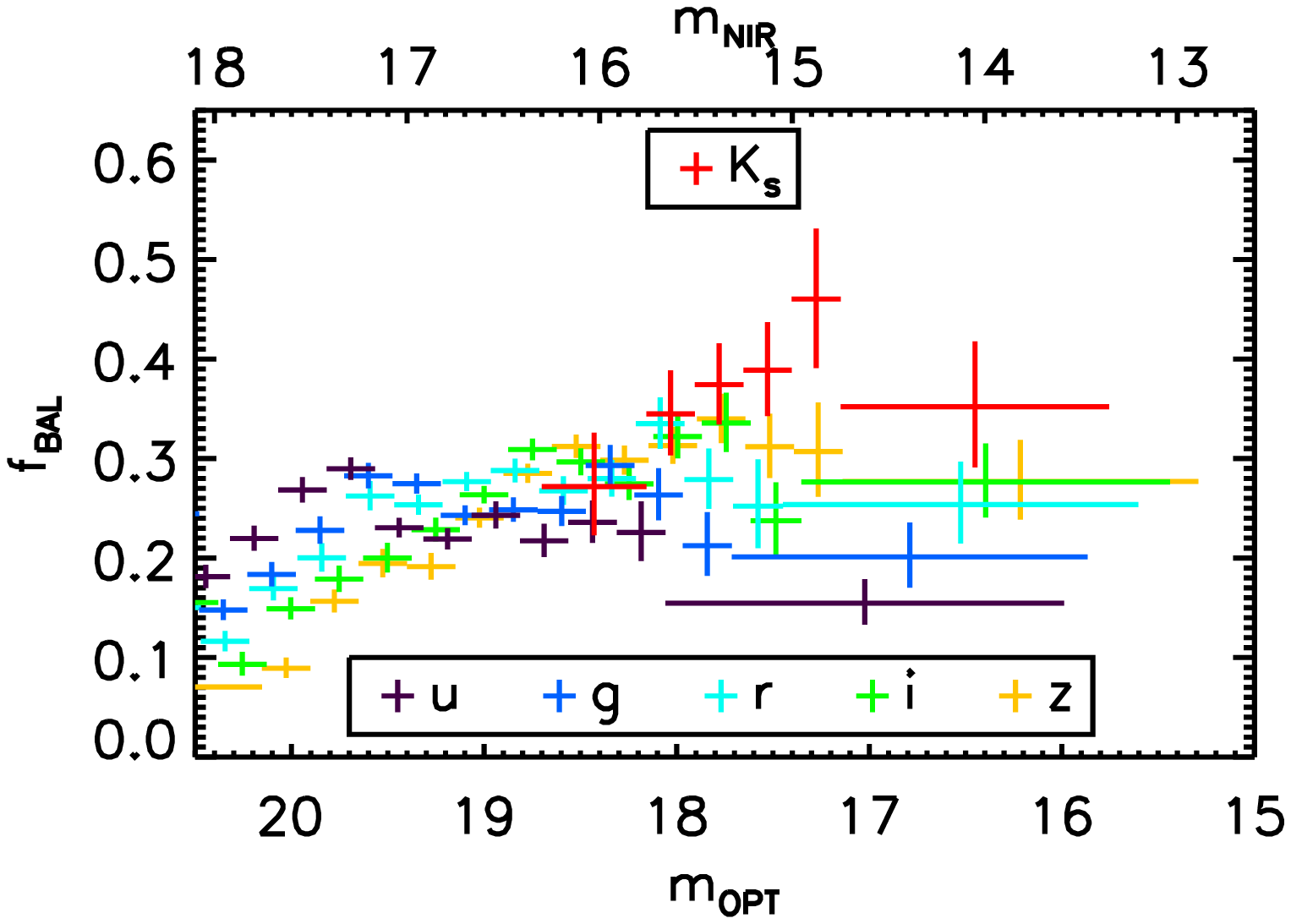}
\caption{
Fractions of BALQSOs as functions of apparent magnitudes in the redshift range of $1.7 < z < 4.38$ ({\it top})
and $1.7 < z < 2.5$ ({\it bottom}). 
 The BALQSO fractions in the $K_s$ band are consistently higher than the optical fractions.
The fractions increase with the wavelengths of the bands. 
\label{fig:fmag}}
\end{figure}

\begin{figure}
\epsscale{0.8}
\plotone{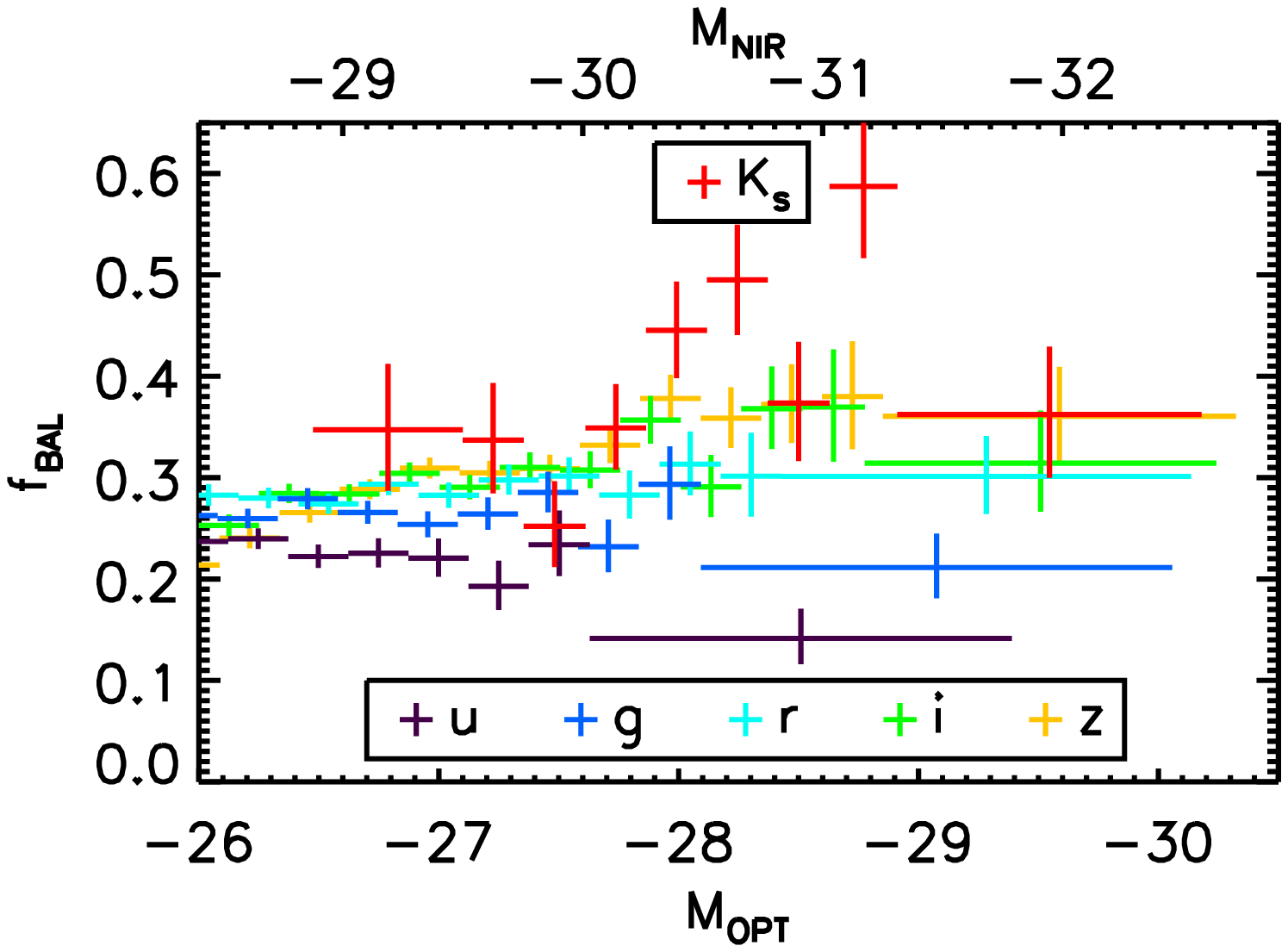}\\
\plotone{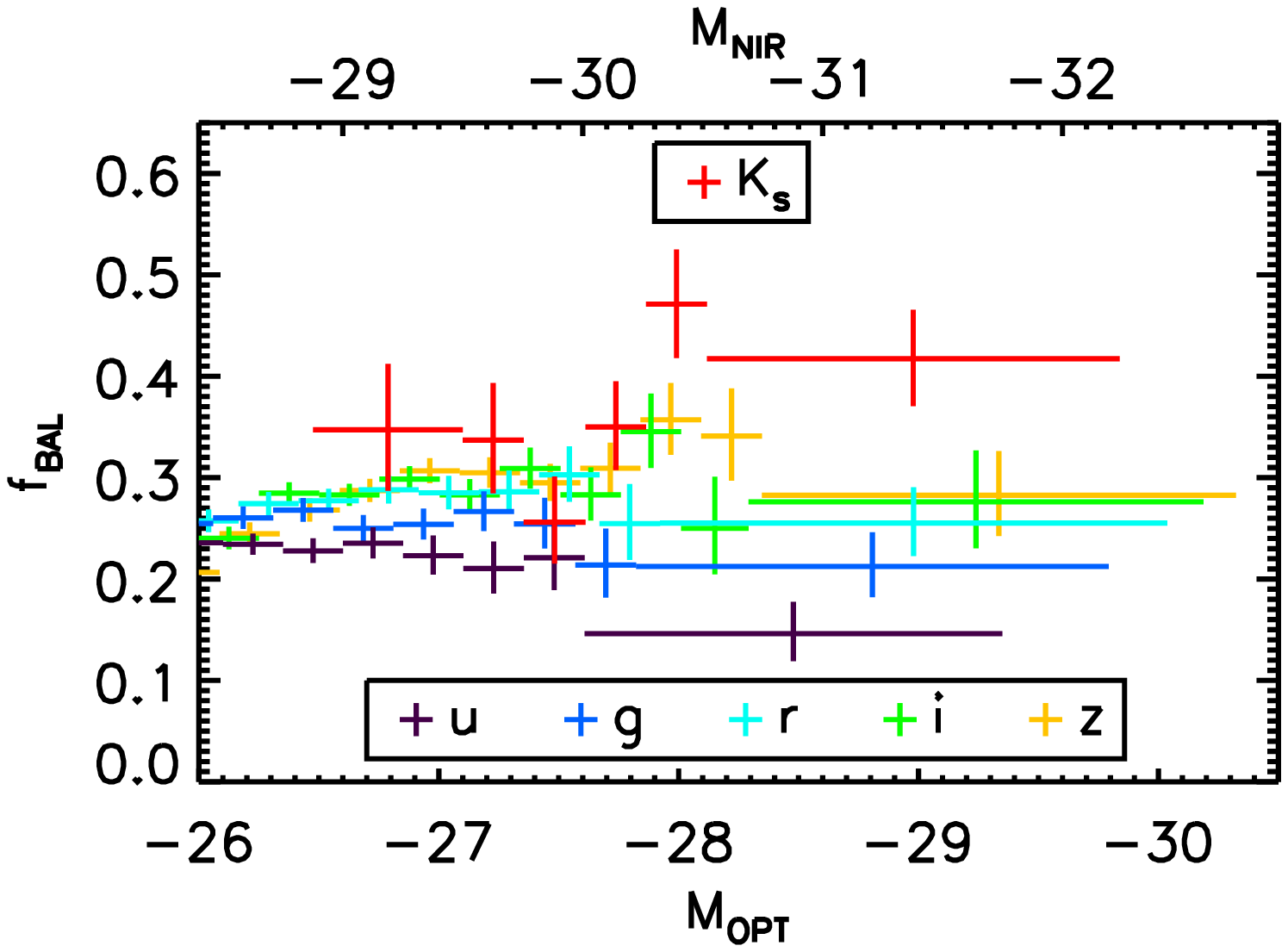}
\caption{
The fractions of BALQSOs as functions of absolute magnitudes in the redshift range of $1.7 < z < 4.38$ ({\it top})
and $1.7 < z < 2.5$ ({\it bottom}). 
As in Fig.~\ref{fig:fmag}, the fractions in the $K_s$ band are consistently higher than the SDSS fractions, and
the fractions increase with the wavelengths of the bands. 
\label{fig:fabs}}
\end{figure}

In top panels of Fig.~\ref{fig:fmag} and Fig.~\ref{fig:fabs}, we plot the fraction of BALQSOs 
against the apparent and absolute magnitudes in the $u$, $g$, $r$, $i$, $z$ and $K_s$ 
bands , respectively, in the redshift range of $1.7 < z < 4.38$.
In the bottom panels of Fig.~\ref{fig:fmag} and Fig.~\ref{fig:fabs}, we also show the same plots 
in the narrower redshift range of $1.7 < z < 2.5$.
The $J$ and $H$ band fractions are very similar to but slightly below the $K_s$ band fractions, 
and we do not show them for clarity reasons.
The K-corrections of the quasars are calculated assuming a power-law spectral index of $\alpha = -0.5$ ($f_\nu \propto \nu^\alpha$, Vanden Berk et al. 2001).
We did not correct the intrinsic obscuration from dust extinction and absorption troughs for BALQSOs.
In both Fig.~\ref{fig:fmag} and Fig.~\ref{fig:fabs}, the BALQSO fraction in the $K_s$ band 
is consistently higher than the optical fractions.
In addition, the optical fractions increase with increasing wavelength.
The two sets of plots are consistent, except for larger 
fluctuations in the $K_s$ band fraction for the full redshift sample, which do not affect our main result.  
In Fig.~\ref{fig:fmag}, we note that the optical fractions drop significantly
when crossing certain magnitude limits.
For example, the $i$ band fraction is roughly a constant of $\sim$0.3 for $i < 19$ mag, 
and steadily decreases as it crosses the $i=19.1$ mag limit.
The $u$, $g$, $r$, and $z$ band fractions behave similarly.
Their drop at different magnitudes from the $i$ band is expected 
given the quasar color and differing sensitivities of the optical bands.
Although the $i=19.1$ mag limit is also the spectroscopic limit for SDSS quasars,
this limit affects differently for BAL and non-BALQSOs.
We caution that there are more uncertainties below this limit, which
make it difficult to model this effect.  

The higher BALQSO fractions in the 2MASS bands, the increase of BALQSO fractions
as wavelength increases, and the drop of BALQSO fractions when crossing certain apparent
magnitudes in the optical bands suggest significant selection biases against BALQSOs.
This is consistent with the spectral differences between BALQSOs and
non-BALQSOs. By definition, significant rest-frame ultra-violet continuum flux
in BALQSOs is absorbed by the absorption troughs. 
In addition, there are also signatures of dust indicated by the difference between
the continuum emission of the BALQSOs and non-BALQSOs (Sprayberry \& Foltz 1992; 
R03).
The composite
spectra of BALQSOs and non-BALQSOs (R03) differ significantly for rest-frame $\lambda < 2400$\AA. 
For the full redshift range considered, $1.7 \le z \le 4.38$, this spectral difference
will affect all of the SDSS filters. However, this spectral difference only
affects 2MASS bands marginally as results of dust extinction. 
The $J$ band is affected for $z>3.7$ BALQSOs,
which only represent a small subset of 2MASS detected quasars. 
In the narrower redshift range, $1.7 \le z \le 2.5$, the $z$ and redder bands are not
affected by the absorption troughs for the Hi-BALQSOs.
Although there will be absorption troughs for Lo-BALQSOs in these red bands, the fraction
of Lo-BALQSOs is very small in this redshift range (T06).
Since the 2MASS survey is
shallower than the SDSS survey, the 2MASS sample is selecting brighter quasars. 
The difference of the BALQSO fractions between 2MASS and SDSS could be explained
if the fractions of BALQSOs are higher for more luminous quasars. 
However,
Fig.~\ref{fig:fabs} shows that the optical fraction is roughly a constant in each
band as functions of absolute magnitudes more luminous than $M_{opt} \sim -26$,
which spans about 5 magnitudes. 
We further
test the idea of optical selection biases by examining the fraction of BALQSOs
using the brightest 245 quasars in each band, such that they have
similar statistics as in the complete $K_s$ sample.
Results are listed in Table~\ref{tab:f}.
The fractions of BALQSOs steadily go up from the $u$ band to the $K_s$ band.
Tests with other sample sizes show consistent trends.

\begin{figure}
\plotone{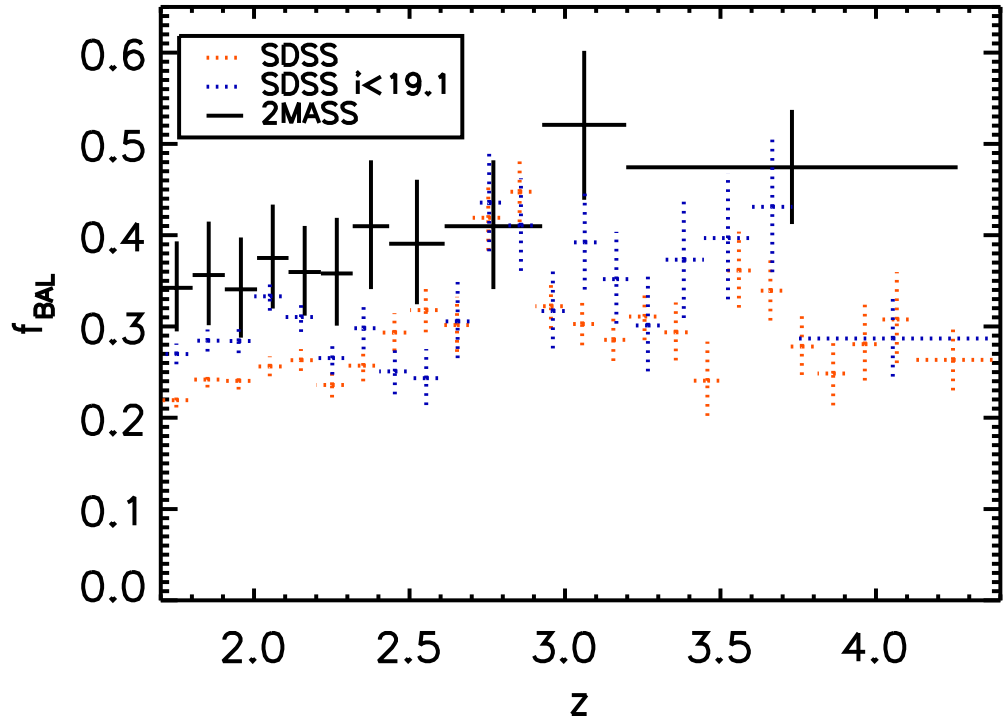}
\caption{
2MASS (solid) and SDSS (dotted) fractions of BALQSOs as functions of redshift. 
The fractions in 2MASS are consistently higher than the SDSS fractions. The
2MASS fractions are less dependent on the redshift than the SDSS fraction,
suggesting less selection biases.\label{fig:fz}}
\end{figure}

Fig.~\ref{fig:fz} shows the fractions of BALQSOs as a function of redshift in the
$K_s$ band for the 2MASS detected quasar sample, and the $i$ band for the SDSS
sample. The 2MASS fractions are consistently higher than the SDSS $i$ band
fractions.
The 2MASS fractions are less dependent on the redshift, while
the optical fractions show significant variations. 
In particular, the optical fractions increase significantly in some redshift bins ($2.7 \le z \le 2.9$ and $z\sim3.7$)
 close to the value expected from 2MASS fractions.
The non-uniformity of the
optical BALQSO fractions as a function of redshift suggests significant selection biases against BALQSOs. 
The excess optical BALQSO fractions at $2.7 \le z \le 2.9$ are
particularly interesting as they are consistent with the near infrared fractions. 
These excess fractions are possibly due to the color difference between BALQSOs and 
non-BALQSOs, which moves BALQSOs further way from the stellar locus.
However, the 2MASS bands do not show significant fraction increases at this redshift range,
which could also be possibly benefited from the fact that the 2MASS filters are 
substantially wider than the SDSS filters.
We also found that this redshift range is possibly related to the gap between SDSS $g$
and $r$ filters, considering a \CIV\ absorption line with a BAL velocity
of $\sim$20,000~\kms.  When the major BAL feature falls between the filter gaps,
there are less differences between BALQSOs and non-BALQSOs.
Thus, the photometric selection
is expected to be less biased against selecting BALQSOs. As a result, the
BALQSO fraction in the redshift range would be unbiased. This effect would also
be important for other gaps between the SDSS filters. We found a similar
increase of BALQSO fraction at the redshift of $z\sim3.7$ corresponding to the
gap between $r$ and $i$ filters. However, we did not find the feature for the
gap between filters $u$ and $g$, which would correspond to the starting edge of
our studied redshift range.  In addition, a blueshift of $\sim$20,000~\kms seems
large given the T06 BALQSO definition.  Detailed modeling of this effect using the SDSS
filter functions, quasar selection algorithm, and composite BALQSO spectrum is needed
to further investigate the origin of the redshift-dependent BALQSO fractions in the
optical bands.

\section{Comparison of 2MASS Properties of BALQSOs and Non-BALQSOs}

\begin{figure}
\plotone{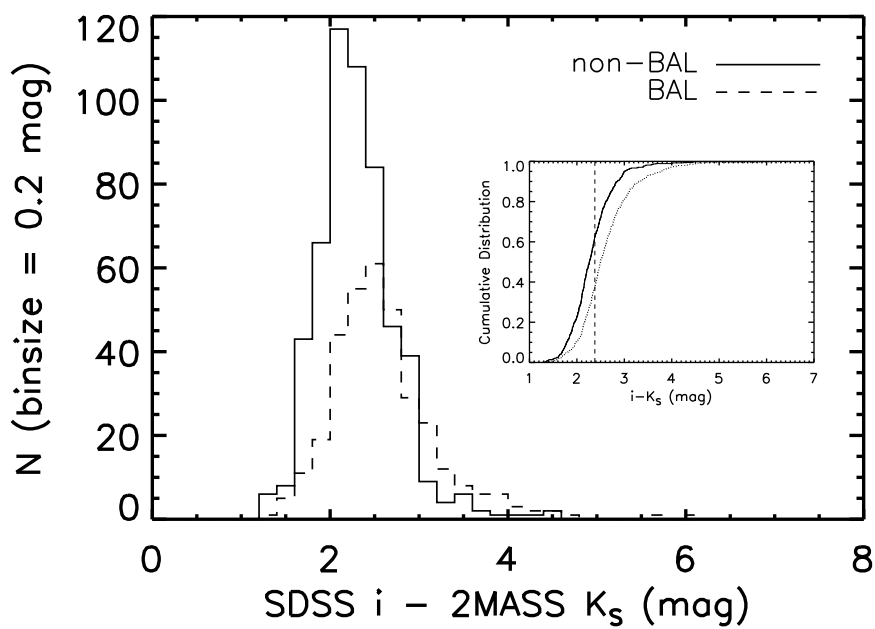}
\caption{The SDSS $i -$ 2MASS $K_s$ color distribution of BALQSOs (dashed line) and non-BALQSOs (solid line) detected in the $J$, $H$, and $K_s$ band in the redshift range of $1.7 \le z \le 4.38$.  The histograms are binned with bin sizes of 0.2 mag.  The inset shows the cumulative distribution of the two samples.  The BALQSOs are redder than the non-BALQSOs and the K-S test results show that the two distributions are significantly different. \label{fig:color}}
\end{figure}

In Fig.~\ref{fig:ik}, we can already see color differences between BALQSOs and
non-BALQSOs. 
The median $i-K_s$ color for BALQSOs and non-BALQSOs are 2.50 and 2.26 mag, respectively,
in the 2MASS sample.
We plot the histograms of the $i - K_s$ color of
BALQSOs and non-BALQSOs of the 2MASS sample in Fig.~\ref{fig:color}.
BALQSOs are redder than non-BALQSOs.  For example, the red tail of the BALQSO
distribution is above that for non-BALQSOs. We performed a K-S test to the two
distributions and obtained a probability of $2\times10^{-12}$ for the null
assumption that the two distributions are drawn from the same parent
distribution.
The same test using only the $K_s$ complete sample yielded a
probability of 0.0001.
This is consistent with our analysis in \S3 that optical
obscuration of BALQSOs can cause significant selection biases against selecting
BALQSOs. The difference of the $i-K_s$ color difference is mainly caused by the $i$ mag
distribution difference between the two populations. 
In the complete $K_s$ sample, the K-S probability for the $i$ mag distributions
of BALQSOs and non-BALQSOs being the same is $4\times10^{-6}$, while the
K-S probability for $K_s$ mag distributions is 0.46.
This result is also consistent with previous studies of BALQSOs
(R03; T06), where BALQSOs are found to be redder in the SDSS bands.
With only nine optically selected BALQSOs, Hall et al. (1997) found no
discrepancy between $B-K$ colors of BALQSOs and non-BALQSOs; however, 
two of their radio selected BALQSOs are particularly red.

The color differences between BALQSOs and non-BALQSOs in the 2MASS bands are smaller
and consistent with the expectation that there is an extra dust extinction in BALQSOs.
The median $J-K_s$ and $H-K_s$ colors are 1.17 and 0.64 mag for non-BALQSOs and 1.26 and 0.67
mag for BALQSOs, with color differences of $0.09\pm0.02$ and $0.03\pm0.02$ mag.
These differences are broadly consistent with the R03 dust extinction fit of $\Delta E(B-V) =0.023$ between BALQSOs and non-BALQSOs, given the observed redshift distribution of the quasars.
We have tested the K-S probability that the near infrared color distributions of
BALQSOs and non-BALQSOs are drawn from the same sample, and found 0.0009 and
0.49 for the $J-K_s$ color distributions in the 2MASS sample and $K_s$ complete sample, 
respectively, and
probabilities of 0.36 and 0.78 for the $H-K_s$ color
distributions.  The K-S test results are consistent with small 2MASS color differences
between BALQSOs and non-BALQSOs.

\section{Simulations}

We performed simple simulations to model the fractions of BALQSOs in the 
optical and near infrared bands.  We show that the increase of the BALQSO fractions
with wavelength can be explained by the selection effects caused by the spectral 
differences between BALQSOs and non-BALQSOs including both absorption lines and
dust extinction.  Our simulations are performed in the redshift range of $1.7 < z < 2.5$.

We assumed that the intrinsic (extinction and absorption corrected) luminosity functions of 
 BALQSOs, $\Phi_{BAL, intr}$, non-BALQSOs, $\Phi_{non-BAL, intr}$, and total quasars, 
$\Phi_{QSO, intr} = \Phi_{BAL, intr}+ \Phi_{non-BAL, intr}$, share the same shape, 
but differ by normalization factors, $\Phi_{BAL, intr} = f_{BAL} \Phi_{QSO, intr}$, 
where $f_{BAL}$ is the intrinsic fraction of BALQSOs.
The shape of the luminosity functions 
are represented by a double power-law luminosity function described by the 
Richards et al. (2005) luminosity function with 
a bright end slope of $\alpha = -3.31$, a faint end slope of $\beta=-1.45$, 
and a break at $M^*(z)=M^*(0)-2.5(k_1z+k_2z^2)$ with $M^*(z=0) = -21.61$, 
$k_1=1.39$, and $k_2=-0.29$ in the $g$ band.
The faint end slope and the break of the luminosity function do not affect our results too much because
the quasars in the sample are all significantly more luminous than the break of the 
luminosity function.  The quasar luminosity 
functions in other bands are obtained by shifting the $g$ band luminosity function 
with the mean color of the quasars (non-BALQSOs) between the $g$ and the other bands.
Since there are extra intrinsic absorption troughs and dust extinction in BALQSOs,
we modeled the obscured luminosity function of BALQSOs, $\Phi_{BAL, obsc}$, by shifting the intrinsic 
BALQSO luminosity function by $\Delta M$ to model the spectral difference between BALQSO and non-BALQSOs, 
i.e., $\Phi_{BAL, obsc}(M+\Delta M)= \Phi_{BAL, intr}(M) = f_{BAL}\Phi_{QSO}(M)$.
  The $\Delta M$ is a function
of wavelength.  
We obtained the values of $\Delta M$ from the composite spectra of BALQSOs and non-BALQSOs (R03).
Besides the absorption troughs, R03 found that an extra SMC dust extinction for BALQSOs of
$\Delta E(B-V) = 0.023$ with $R_V = 2.98$ fits the continuum spectral differences.
We therefore modeled the $\Delta M$ as two components, with one caused by such a dust extinction
and the other by absorption troughs.
Since the $u$ band is not completely covered 
by the composite
spectra of R03, we cannot fully model the absorption trough dimming and therefore 
exclude the $u$ band
in our fitting process.  The $z$, $J$, $H$, and $K_s$ bands are also not completely covered; 
however, since there are few absorption troughs in these band for quasars with $1.7 < z < 2.5$, 
we modeled the $\Delta M$ from dust extinction only.
We list the values of $\Delta M$ in Table~\ref{tab:dm}.

\begin{figure}
\plotone{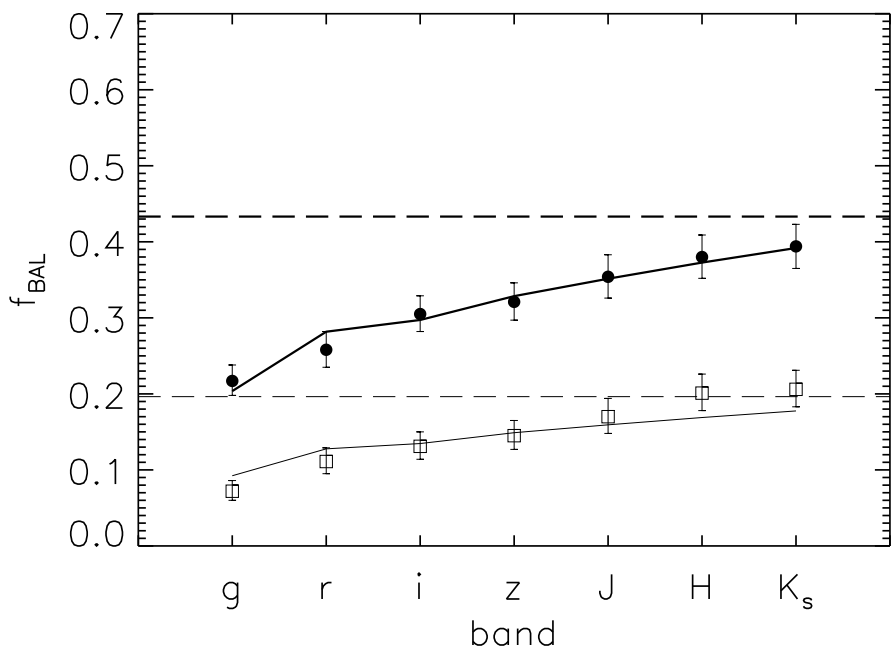}
\caption{The BALQSO fractions in the $g, r, i, z, J, H$, and $K_s$ bands using T06 definition (filled circles),
and the fitting results from our simulations (thick solid line).
The thick dashed line show the true fraction of BALQSOs, $f_{BAL} = 0.43\pm0.02$, obtained from the simulation.  We also show the BALQSO fraction (squares) using Weymann et al. (1991) definition, and the corresponding simulation results (thin solid line) and the true BALQSO fraction $f_{BAL} = 0.20\pm0.02$ (thin dashed line). \label{fig:model}}
\end{figure}

 As discussed earlier, we performed our simulations in the narrower redshift bin $1.7 < z < 2.5$.
This range avoids the larger optical BALQSO fractions at $2.7 < z < 2.9$
and $z\sim3.7$.  We examined the 2MASS detected quasar absolute
magnitudes as a function of redshift.  In our redshift range, 2MASS
detects quasars more luminous than $M_{K_s} < -30.85$ and $-30.1$~mag
for the $K_s$ complete sample and 2MASS sample, respectively. To
translate these limits to the other bands, we applied the non-BALQSO
colors to the magnitude limits. For each band and limit, we calculated
the BALQSO fractions and present them in Table~1. 
Both limits agree
qualitatively, and we hereafter use the larger 2MASS sample.
In particular, the $g$, $r$, $i$, $z$, $J$, $H$, and $K_s$ limits are -27.46, -27.62, -27.75, -27.91, -28.86, -29.41, and -30.10 mag, respectively.
We started with quasar luminosity functions at the mean redshift, $z=2.1$, of our narrow redshift sample
and considered the fractions from the $g$ to $K_s$ bands.
We fit these BALQSO fractions using our luminosity functions for BALQSOs and non-BALQSOs with one free parameter, $f_{BAL}$, and obtained $f_{BAL} = 0.43\pm0.02$
with $\chi^2 = 1.62$ for 6 degrees of freedom (Figure~\ref{fig:model}).
We tested the robustness of our result by using the quasar luminosity functions at $z=1.7$ and $z=2.5$ in our fitting process, and obtained
consistent fitting results.
The fitting results are insensitive to the bright end slope of the luminosity function within the error of $\Delta \alpha \sim 0.05$ (Croom et al. 2004).

Alternatively, it is possible that BALQSOs are more near infrared luminous, which might
explain the higher BALQSO fractions in the $J$, $H$, and $K_s$ bands.
We tested this idea by fixing $f_{BAL}$ with the value from optical fractions and applying an extra brightening
of $\Delta M_{NIR}$ for BALQSOs compared to the previous simulation.
We found $\Delta M_{NIR} \sim 0.2$~mag is needed to fit the $J$, $H$, and $K_s$ band
fractions.  However, the fractions of the $g$, $r$, and $i$ bands from this model are significantly
lower than the observations.  
Given our knowledge of the dust extinction and absorption troughs in BALQSOs, a wavelength dependent brightening of 
BALQSOs is needed from the $g$ to $K_s$ band to fit the observed fractions with a low intrinsic BALQSO fraction
of 25--30\%.
Therefore, we consider this model as highly fine-tuned.
This supported by the recently result that there is no significant spectral difference between BALQSOs and non-BALQSOs in mid-infrared within a moderate sized sample (Gallagher et al.\ 2007).

\section{Summary and Discussion}
We have presented enhanced fractions of BALQSOs in the 2MASS detected quasars
within the SDSS DR3 sample compared to fractions derived in optical bands.
The relatively large number of near infrared luminous BALQSOs used in our study
compared with previous studies argues against discrepancies from small number
statistics.
The BALQSO fraction is  $40.4^{+3.4}_{-3.3}$\% in the $K_s$
complete sample and $38.5^{+1.7}_{-1.7}$\% in the 2MASS sample.
The complete $J$ and $H$ samples also have similar fractions. 
After plotting the BALQSO fractions against apparent and absolute magnitudes
in the 2MASS and SDSS bands, we found the 2MASS fractions are consistently
higher than the optical fractions.
In addition, the BALQSO fractions steadily increase with wavelength from the
SDSS $u$ to the 2MASS $K_s$ bands.
We conclude that the optical photometric systems have significant
selection biases against BALQSOs, and the BALQSO
fraction in the near infrared more accurately reflects the true fraction of BALQSOs. 
This is also supported by the $i - K_s$ color difference between the populations
of BALQSOs and non-BALQSOs.

After using a simple simulation incorporating the luminosity function of quasars
and the amount of obscuration for BALQSOs, we are able to simultaneously fit the BALQSO fractions 
in the SDSS and 2MASS bands and obtain a true BALQSO fraction of $43\pm2$\%,
significantly higher than other BALQSO fractions reported in the literature.
Although a small sample was used, Chartas (2000) found a BALQSO fraction of 35\%
in a gravitationally-lensed quasar sample, close to our near infrared BALQSO
fraction.
This sample actually avoids some of the selection biases in the optical bands. 
The previous lens sample was selected serendipitously from many surveys, and all
the gravitational lenses are identified spectroscopically. In addition, the
lensing has boosted the S/N of the quasars so that they are less biased against
discounting the BALQSOs.

Selection biases against BALQSOs has been proposed by previous studies,
where a continuum anisotropy causes the selection biases (e.g., Goodrich 1997; Krolik \& Voit 1998).
In our simulations we show that the selection biases can be explained with 
 obscuration from dust extinction and absorption troughs in BALQSOs.
Essentially, an anisotropy is also produced after the optical obscuration
which, combined with the steep quasar luminosity function, will produce lower
BALQSO fractions in those bands more sensitive to obscuration.
An intrinsic continuum anisotropy is also possible; however, it needs to be fine tuned
as a function of wavelength that behaves similarly to the effect of dust extinction,
especially in the longer wavelength bands.
Different continuum emission between BALQSO and non-BALQSOs is also expected if the 
BALQSOs are
either at special evolution stages of quasar evolution, or the near infrared emission
of BALQSOs are enhanced by the reprocessed emission from the absorption lines. 
This fact is also connected to the idea that BALQSOs are in the transition stage
between Infrared Luminous Galaxies and quasar phases (L{\'{\i}}pari et al. 
2005). 
To test whether the continuum emission in the BALQSOs and non-BALQSOs are
similar, we need to compare the absorption corrected spectra for BALQSOs with
non-BALQSOs, which involves significant complexity in the analysis of the
optical spectra. As the observed near infrared band is little affected by the BAL
features, it would be ideal to make the comparison there. 
However, a near infrared spectroscopic survey is needed to better measure the 
power-law slope and K-correction of the quasars before we draw a solid conclusion.
Our sample size is limited by the 2MASS survey limits.  Future infrared surveys
(e.g., UKIDSS with survey limits of $K = 18.3$~mag) will significantly increase
the sample size, and enable us to extend this study of BALQSO fractions to less luminous quasars.

Our analysis indicates that the near infrared BALQSO fraction more accurately 
reflects the true fraction of BALQSOs.
This is important when using the fraction of BALQSOs to constrain quasar geometric
and evolutionary models.  In this paper, we found a
BALQSO fraction of $\sim40$\% in the 2MASS bands under the T06 criterion. This
indicates a correction of factor $\sim1.5$ is needed for the optical fractions,
consistent with the modeling of Hewett \& Foltz (2003). 

We note that the fraction of BALQSOs is also
dependent on the spectral definition of BALQSOs. T06 also analyzed the fraction
of BALQSOs under the original definition of Weymann et al. (1991) and found
a BALQSO fraction of 10\%, different from the 26\% under the T06 criterion.
We analyzed the fraction of BALQSOs in 2MASS under the Weymann et al. (1991) definition
and obtained fractions of $23\pm3$\% and $20\pm2$\% in the $K_s$ complete and 2MASS
sample using the balnicity indices provided by T06.
We obtained a correction factor of $\sim 2$ for the optical fraction under the 
Weymann et al.\ definition.
This result is not surprising since the Weymann et al.\ definition 
for BALQSOs is more strict, and we expect more severe optical obscuration and
a larger correction factor for the optical fraction. 
This fraction is also very similar to $22\pm4$\% obtained by 
modeling directly in the optical bands and using the Weymann et al.\ definition (Hewett \& Foltz 2003).

Since the radio bands are also not affected by dust extinction and absorption troughs,
we expect the radio fraction of BALQSOs should also have a large fraction of $\sim40$\%
under the T06 definition.
However, the situation in this case is more complex.
For example, the fractions of BALQSOs are found to be dependent on the radio flux (Hewett \& Foltz 2003),
while our BALQSO fractions in the near infrared bands are nearly constant, independent of the flux.
The complexity may be due to different mechanisms responsible for the radio and UV/optical (observed optical/near infrared)
emission of quasars;
however, further investigation is needed.

Our near infrared BALQSO fractions are
roughly constants across the large redshift range of $1.7 < z < 4.38$, which suggests
that the covering fraction of the BAL wind does not evolve
significantly in the geometric models. 
The 40\% BALQSO fraction implies a large wind half opening angle of
$\sim 24^{\circ}$, suggesting that the BAL wind is raised significantly above the accretion disk
as compared to $\sim 6^{\circ}$ for a $\sim 10$\% fraction.
This might impose significant challenges to theoretical models on wind dynamics (see Krolik \& Voit 1998 for this and other
implications). 
The results of the paper will also be useful in constraining evolutionary
models for BALQSOs by combining the constraints
from the number density, luminosity function, and duty cycle of quasars.

\acknowledgements
We acknowledge discussion at the AGN lunch in the OSU astronomy department, especially Karen Leighly, Chris Kochanek, David Weinberg, and Paul Martini. We thank the anonymous referee for helpful suggestions.

\clearpage

\begin{deluxetable}{ccccccccc}
\tabletypesize{\scriptsize}
\tablecolumns{9}
\tablewidth{0pt}
\tablecaption{Fraction of BALQSOs in SDSS and 2MASS Bands \label{tab:f}}
\tablehead{
\colhead{Sample\tablenotemark{a}} &
\colhead{$u$} &
\colhead{$g$} &
\colhead{$r$} &
\colhead{$i$} &
\colhead{$z$} &
\colhead{$J$} &
\colhead{$H$} &
\colhead{$K_s$} 
}
\startdata
Brightest 245 & $15.1^{+2.6}_{-2.3}$\% & $19.6^{+2.9}_{-2.6}$\% & $24.9^{+3.1}_{-2.9}$\% & $26.9^{+3.1}_{-2.9}$\% & $29.4^{+3.2}_{-3.0}$\% & $36.3^{+3.3}_{-3.2}$\% & $38.4^{+3.3}_{-3.3}$\% & $40.4^{+3.4}_{-3.3}$\% \\
Brightest 884 & $20.1^{+1.4}_{-1.4}$\% & $23.9^{+1.5}_{-1.5}$\% & $28.2^{+1.6}_{-1.5}$\% & $30.7^{+1.6}_{-1.6}$\% & $32.2^{+1.6}_{-1.6}$\% & $36.4^{+1.7}_{-1.7}$\% & $38.7^{+1.7}_{-1.7}$\% & $39.5^{+1.7}_{-1.7}$\%\\
Luminous $K_s$ complete & $14.8^{+3.4}_{-2.9}$\% & $22.2^{+6.5}_{-5.5}$\% & $28.3^{+8.2}_{-7.1}$\% & $29.8^{+7.3}_{-6.5}$\% & $29.6^{+7.5}_{-6.6}$\% & $36.7^{+8.1}_{-7.5}$\% & $34.5^{+7.6}_{-6.9}$\% & $32.7^{+7.8}_{-7.0}$\%\\
Luminous 2MASS\tablenotemark{b} & $20.3^{+1.3}_{-1.3}$\% & $21.7^{+2.1}_{-1.9}$\% & $25.8^{+2.4}_{-2.3}$\% & $30.5^{+2.4}_{-2.3}$\% & $32.1^{+2.5}_{-2.4}$\% & $35.4^{+2.9}_{-2.8}$\% & $38.0^{+2.9}_{-2.8}$\% & $39.4^{+2.9}_{-2.9}$\%\\
\enddata
\tablecomments{The increasing BALQSO fractions from $u$ to 
$K_s$ band showing significant wavelength dependence.}
\tablenotetext{a} {The first two lines list the fractions of BALQSOs calculated from samples using the brightest N quasars in each bands. The numbers 245 and 884 are used to match the numbers of
quasars in the $K_s$ complete sample and the 2MASS sample. 
The bottom two lines list the fractions of BALQSOs in $1.7 < z < 2.5$ calculated above absolute magnitudes 
 for the $K_s$ complete (-27.66, -28.21, -28.37, -28.50, -28.66, -29.61, -30.16, and -30.85 mag) sample and the 2MASS sample (-26.91, -27.46, -27.62, -27.75, -27.91, -28.86, -29.41, and -30.10 mag) from the $u$ to $K_s$ bands, respectively.
In the redshift range of $1.7 < z < 2.5$, 2MASS
detects quasars more luminous than $M_{K_s} < -30.85$ and $-30.1$~mag
for the $K_s$ complete sample and 2MASS sample, respectively. To
translate these limits to the other bands, we applied the non-BALQSO
colors to obtain the magnitude limits in other bands.}
\tablenotetext{b}{These values are used in \S5 and Figure~6.}
\end{deluxetable}

\clearpage

\begin{deluxetable}{ccccccccc}
\tabletypesize{\scriptsize}
\tablecolumns{8}
\tablewidth{0pt}
\tablecaption{Obscuration Magnitudes of BALQSOs Due To Dust Extinction and Absorption Troughs Compared to Non-BALQSOs \label{tab:dm}}
\tablehead{
\colhead{$\Delta M$ From} &
\colhead{$g$} &
\colhead{$r$} &
\colhead{$i$} &
\colhead{$z$} &
\colhead{$J$} &
\colhead{$H$} &
\colhead{$K_s$} 
}
\startdata
Dust Extinction    & 0.28 & 0.20 & 0.16 & 0.13 & 0.10 & 0.07 & 0.05 \\
Absorption Troughs & 0.08 & 0.01 & 0.02 & \nodata &  \nodata & \nodata & \nodata \\
Total              & 0.36 & 0.21 & 0.18 & 0.13 & 0.10 & 0.07 & 0.05 \\
\enddata
\tablecomments{The dust extinction magnitudes are calculated based on the continuum spectral differences of $\Delta E(B-V) = 0.023$ with $R_V = 2.98$ (Reichard et al.\ 2003).  The $\Delta M$ from absorption troughs are calculated from the difference between the composite spectra of BALQSOs and non-BALQSOs (Reichard et al.\ 2003) excluding dust extinction assuming $z\sim2.1$.
Specifically, we first de-redden the Reichard et al. BALQSO composite
spectrum by E(B-V)=0.023 and then convolve the de-reddened spectrum with the SDSS filter
functions to determine the $\Delta M$ for the absorption troughs.}
\end{deluxetable}

\clearpage

\clearpage

\clearpage

\clearpage

\clearpage

\clearpage

\end{document}